\documentstyle[nato,psfig]{crckapb}

\begin{opening}
\title{OLD OPEN CLUSTERS AS TRACERS OF GALACTIC \protect \\
EVOLUTION}

\author{A. BRAGAGLIA}
\author{M. TOSI}
\institute{Osservatorio Astronomico di Bologna, via Ranzani 1\\
      I-40127 Bologna (Italy), e-mail: angela, tosi@bo.astro.it}
\author{G. MARCONI$^{1,2}$}
\institute{$^1$ Osservatorio Astronomico di Roma, Via Osservatorio 5\\
      I-00040 Monte Porzio (Italy), \\
           $^2$ E.S.O., Santiago (Chile), e-mail: gmarconi@eso.org}
\author{E. CARRETTA}
\institute{Osservatorio Astronomico di Padova, vicolo Osservatorio 5\\
      I-35122 Padova (Italy), e-mail: carretta@pd.astro.it}

\end{opening}

\runningtitle{Old Open Clusters}

\begin{document}

\section{Why (old) open clusters?}
Open clusters (OC's) are important to study the properties of the Galactic
disk, since they offer information on ages, both absolute and relative, and on
the metallicity evolution, both in space and time. In fact, OC's are found in
different regions of the disk, cover a large interval in age (from a few Myr
to about 10 Gyr) and in metallicity (Z=Z$_\odot$/20 to supersolar). Moreover,
their ages and distances are accurate, much more than for any other disk
object like e.g., single field stars. Also, since OC's orbits do not generally
take them too far away from their birthplaces, we may assume that their
current position in the Galaxy is representative also of their original one,
and that we are not smearing properties by mixing populations, always a
problem when dealing with field stars.

To define at least a reliable ranking of the open clusters properties we need
a large sample of objects whose age, distance and  metallicity are accurately
and homogenously known: see e.g., Janes \& Phelps (1994), Carraro \& Chiosi
(1994), Friel (1995), Twarog et al. (1997). These authors have tried to define
such samples, but had to sacrifice something in the quality of the data, often
taken from the literature, and not as excellent as attainable with today's
means, and/or on the homogeneity of the treatment to get a sample as large as
possible.

We have started a few years ago a project, admittedly ambitious, to build a
homogeneous and statistically significant sample of clusters of various ages,
metallicities and positions in the disk. We work only on new photometry taken
by our group, or on literature data of comparable quality. The clusters'
properties are derived using the synthetic colour-magnitude diagram  method
(Tosi et al. 1991). We compare the observed diagram to a grid of synthetic
ones, generated from a series of homogeneous sets of theoretical evolutionary
tracks by several authors. The comparison is based both on morphology (e.g.,
main sequence shape, red clump position, gaps, etc.) and on population ratios
(e.g., the luminosity function). We have already applied this method to nine
open clusters, ranging in age from 0.1 to about 10 Gyr (see Bragaglia et al.
1999 for a recent review), and we have data for ten more.

We plan now to turn to spectroscopy, in particular to high resolution, in order
to get precise information on the clusters' metallicity.

\begin{figure}
\vspace{15cm}
\caption{Old open clusters: run with Galactocentric distance of the observed
metallicity, derived with different techniques (from top to bottom: by low
resolution spectroscopy, by Washington, UBV, DDO photometry, and by high
dispersion spectroscopy). The slope of the resulting metallicity gradient is
indicated in each panel.}
\includegraphics{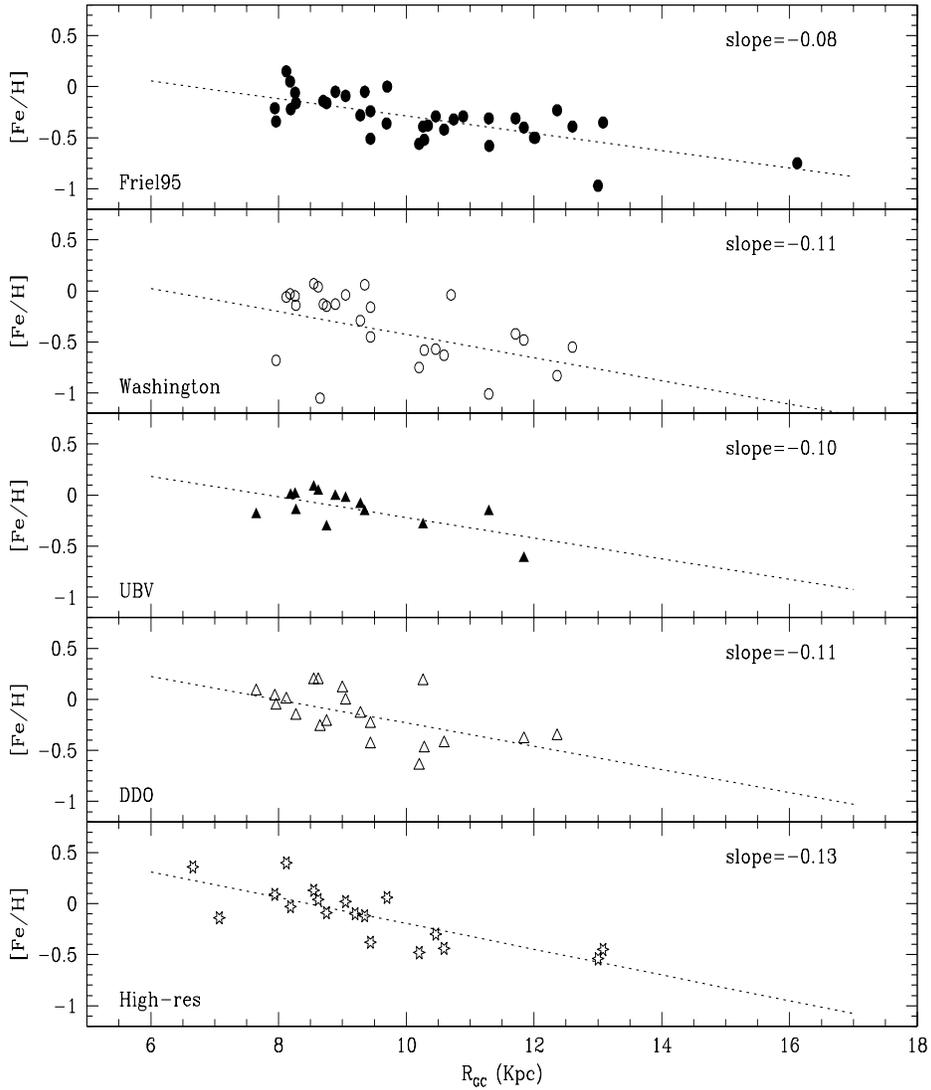}
\end{figure}

\section{What do we know about open clusters' metallicities?} Of all the OC's
we have studied photometrically, four have the metallicity measured by high
dispersion spectroscopy, the best technique to get reliable results, with
uncertainties of less than about 0.1 dex in [Fe/H] (see Gratton 1999 for a
recent review). If we look at the whole sample of old OC's this is true for
only 18 of the about 80 known. There are other ways to get the metallicity,
like low resolution spectroscopy calibrated to high-res, as done by Janes and
Friel (see e.g., Friel 1995; this allows uncertainties of about 0.15 dex), or
Washington, UBV and DDO photometry (with uncertainties of about 0.2 dex).  A
little more than half of the old OC's sample has the metal abundance measured
by at least one of the above methods. An interesting feature comes out of the
comparison between metallicities measures by high-res spectroscopy and any of
the other methods: they all underestimate [Fe/H] by at least 0.1 dex on
average (even if the individual values may be off by much more).

Apart from a possible sistematic effect, does the technique used influence the
derived general properties of the cluster sample? If we consider the old OC's
we can study the existence of the radial metallicity gradient using
metallicities derived by all methods. Figure 1 shows the different cases; at
least part of the observed dispersion at all Galactocentric distances is real.
If we determine the slope of the metallicity gradient we find values very
similar to each other within the errors; they are indicated in Figure 1, and
are of the order of --0.1 dex kpc$^{-1}$.
Notice the larger slope obtained for the high-res values; the sample is still
too small to decide whether it is really significant, and it would be
interesting to have many more high-res determinations to put on firmer grounds
the absolute and relative values of OC's metallicities.

Regarding the slope of the radial gradient, it seems irrelevant which method
is used to determine metallicities. On this basis, to study whether the
metallicity gradient has changed with time, we use the Friel (1995) values,
since they are the most numerous for the old OC's (36 OC's, compared to 26
with Washington, 14 with UBV and 20 with DDO photometry). If we divide the
whole sample in four age intervals (age $<$ 1, 1--3, 3--6, $>$ 6 Gyr) we find
that there is no variation of the gradient slope with age (see Figure 2).
There may be a slight indication of a shallower slope for the younger OC's,
but it's well within the errors.

Finally, using samples with metallicities obtained homogenously, we do not see
any indication of the discontinuity at R$_{\rm GC}$ about 10 kpc assumed by
Twarog et al. (1997) as an indication of the edge of the original thick disk.

\begin{figure}
\vspace{13cm}
\caption{Run with Galactocentric distance of the observed metallicity (as
derived by Friel 1995) for open clusters, divided into four age bins.}
\includegraphics{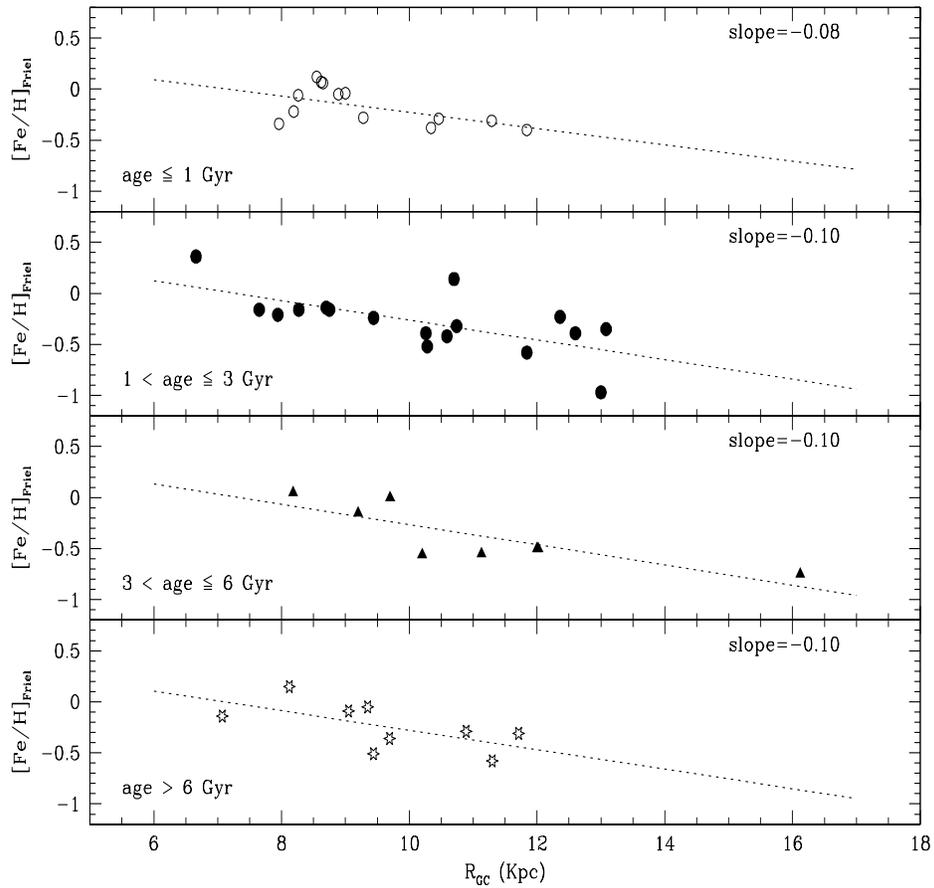}
\end{figure}

\section{The metallicity of NGC 6253 and future work}
Our first direct contribution to the knowledge of OC's metallicities comes
from NGC 6253, a cluster in the direction of the Galactic centre, with age
$\simeq $ 3 Gyr, and metallicity solar or (more probably) twice solar as
derived from photometry alone (Bragaglia et al. 1997). We have high resolution
spectra of four giant stars, taken with EMMI mounted on NTT as part of a
backup program in a night not suitable for photometry. As a result, the
signal-to-noise ratio (about 35) and the resolution (about 15,000) are not as
good as one would desire to derive precise abundances from fine analysis of
high resolution spectra. This is especially true for stars of metallity so
high that continuum tracing is a very difficult task, and for this reason we
compared the observed spectra with a field star of similar metallicity
($\zeta$ Cygni, [Fe/H]=+0.05), as shown in Figure 3.

On the basis of their radial velocities, all the four stars are cluster
members. We have only studied the iron content of two red clump stars (see
Carretta et al. 1999 for a more detailed description of the followed
procedure), and derived for them the following values: \#2971: [Fe/H] = +0.33,
and \#2508: [Fe/H]= +0.39. The overall uncertainty, resulting from the
measured equivalent widths and the adopted atmospheric parameters, is
0.15--0.20 dex, so we may assume for NGC 6253: [Fe/H] = +0.36 $\pm$ 0.20 dex.
This is in very good agreement with the photometric study and with the value
found by Piatti et al. (1998) from integrated spectra.
NGC 6253 appears in Figure 1 (high resolution panel) as the innermost cluster
of the whole sample, and its metallicity conforms to the Galactocentric
gradient found for the other clusters.

We have recently acquired spectra of better quality of the same stars in NGC
6253, and we plan to analyze them to derive the abundances of other elements.
Our plan for the next future is to obtain high-resolution spectra in several
open clusters in order to derive accurate and homogenous metallicities of a
large sample of clusters. This is a prerequisite for a really meaningful study
of the history of the chemical enrichment of the Galactic disk.

\begin{figure}
\vspace{8cm}
\caption{A region of the normalized spectra of the two NGC 6253 stars,
compared to $\zeta$ Cyg: notice the shallower lines of the latter, indicative
of a slightly lower metal content; the spectra have been arbitrarily shifted
for clarity.}
\includegraphics{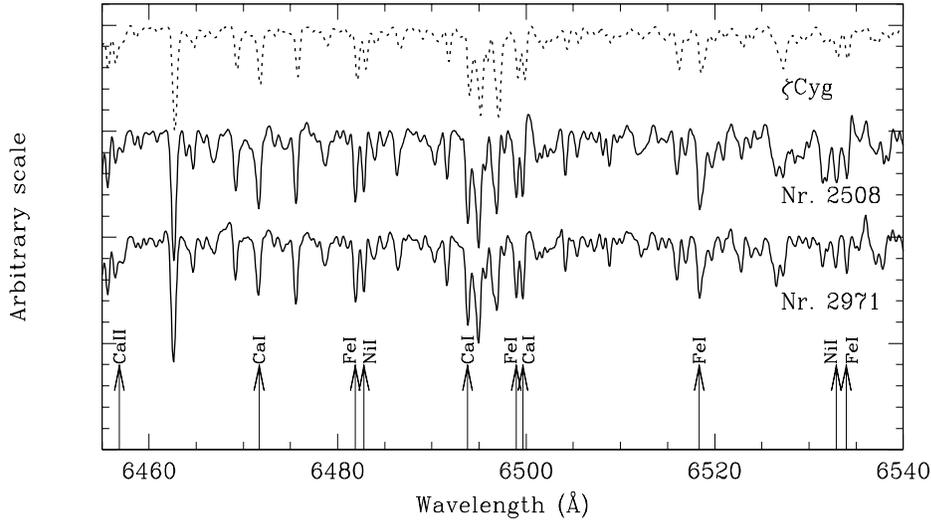}
\end{figure}

{}

\end{document}